\documentclass{aa}  
\usepackage{natbib}
\bibpunct{(}{)}{;}{a}{}{,}
\usepackage[T1]{fontenc}
\usepackage{ae,aecompl}
\usepackage{graphicx}
\usepackage{amsmath}
\usepackage{amssymb}
\begin{document}

   \title{Drifting inwards in protoplanetary discs II: \\The effect of water on sticking properties \\at increasing temperatures}
   
\titlerunning{Drifting inwards in protoplanetary discs II}

   \author{     C. Pillich\inst{1}
   \thanks{E-mail: cynthia.pillich@uni-due.de}
   \and T. Bogdan\inst{2}
        \and
        J. Landers\inst{1}
        \and
        G. Wurm\inst{2}
        \and
        H. Wende\inst{1}
}

\authorrunning{Pillich et al.}

\institute{     University of Duisburg-Essen and Center for Nanointegration Duisburg-Essen (CENIDE), Faculty of Physics, Lotharstr. 1, 47057 Duisburg, Germany\\
        \and
University of Duisburg-Essen, Faculty of Physics,
        Lotharstr. 1, 47057, Germany\\
}

\date{Received xy. Accepted z.}

\abstract{
In previous laboratory experiments, we measured the temperature dependence of sticking forces between micrometer grains of chondritic composition. The data showed a decrease in surface energy by a factor $\sim 5$ with increasing temperature.
Here, we focus on the effect of surface water on grains. Under ambient conditions in the laboratory, multiple water layers are present. At the low pressure of protoplanetary discs and for moderate temperatures, grains likely only hold a monolayer. As dust drifts inwards, even this monolayer eventually evaporates completely in higher temperature regions. 
To account for this, we measured the tensile strength for the same chondritic material as was prepared and measured under normal laboratory conditions in our previous work, but now introducing two new preparation methods: drying dust cylinders in air (dry samples), and heating dust pressed into cylinders in vacuum (super-dry samples).
For all temperatures up to 1000\,K, the data of the dry samples are consistent with a simple increase in the sticking force by a factor of $\sim 10$ over wet samples. 
Up to 900\,K super-dry samples behave like dry samples. However, the sticking forces then exponentially increase up to another factor $\sim 100$ at about 1200\,K. 
The increase in sticking from wet to dry extends a trend that is known for amorphous silicates to multimineral mixtures. The findings for super-dry dust imply that aggregate growth is boosted in a small spatial high-temperature region around 1200\,K, which might be a sweet spot for planetesimal formation.}


\maketitle

\section{Introduction}
\label{sec:introduction}

This work complements the work by \citet{Bogdan2020b} (paper\,I). In this earlier work, we studied the change in sticking properties of chondritic dust because it is subject to increasing temperatures, which alters the grain composition. In the work presented here, we quantify the role of surface water in this process. 

In short, sticking properties of dust-sized grains of different composition decide the early phases of planet formation and how far grains can grow and might further evolve \citep{Youdin2005, Blum2008, Dominik1997, Kataoka2013, Johansen2014, Musiolik2016, Gaertner2017, deBeule2017, Demirci2017, Gundlach2018, Musiolik2019, Bogdan2019, Demirci2019b, Kimura2015,Steinpilz2019,Yamamato2014}. It is an ongoing endeavour. Based on a very general mechanism of sub-Keplerian gas motion, particles drift inwards \citep{Weidenschilling1977}. This drift of solids changes their composition.
This will bring matter from outside the water snowline to the hot inner region \citep{Banzatti2020}. Water or ice will readily sublimate just inside the snowline and likely leave no more than a monolayer on the surfaces of the grains \citep{Kimura2015}. The last of this might yet not be removed from the surface even when 500\,K are reached \citep{Angelo2019}. However, the grains still become warmer the farther inwards they drift and adsorbed water on their surfaces will eventually vanish completely \citep{King2021}.
Therefore the sticking forces of grains likely depend on the bulk composition, as we showed in paper\,I, for instance, but they will be altered by water layers of various thickness.

To approach a realistic mix of minerals and its evolution in protoplanetary discs, we used a chondrite in \citet{Bogdan2020b} that we milled to micrometer dust. This dust was then subjected to increasing temperatures for 1\,h under vacuum.
After each heating step, the dust cooled down to room temperature. We then formed dust cylinders and used a Brazilian test to measure the splitting tensile strength of the dust cylinders. From this, we deduced an effective surface energy for the chondritic dust of $\gamma = 0.07 \rm \, J/m^2$ at room temperature, which is on the order typically used for amorphous $\rm SiO_2$. Compositional changes lead to a decrease in this value by a factor of 5 from room temperature to 1200\,K. From the iron-bearing minerals measured, iron oxide and kamacite have high surface energies. For the wet dust, the ratio of kamacite and iron oxide on the overall composition decreases. Therefore the decrease in surface energy is likely linked to their decline.

In \citet{Bogdan2020b}, we did not take care of any surface water and only traced the effect of compositional changes (plus water). However, based on \citet{Steinpilz2019} and \citet{Kimura2015}, it is known that the surface water strongly alters the sticking forces, that is, water decreases the sticking forces by about a factor of 10 for amorphous silica. Recent measurements of contact forces on meteoritic powder by \citet{Nagaashi2021} also point in this direction.
Our goal in this work is to further quantify the effect of water on the surface of minerals other than amorphous silica, that is, determine how far the composition dominates the surface energy below a water mantle, or by what fraction the surface energy is altered in the transition from no surface water over monolayer to multilayer.

\section{Laboratory experiments}
\label{sec:experiments}

The measurement techniques employed are the same as in \citet{Bogdan2020b} (paper\,I). 
We therefore only highlight the changes and very basic procedures here and refer to paper\,I for details.

\begin{figure}
        \includegraphics[width=\columnwidth]{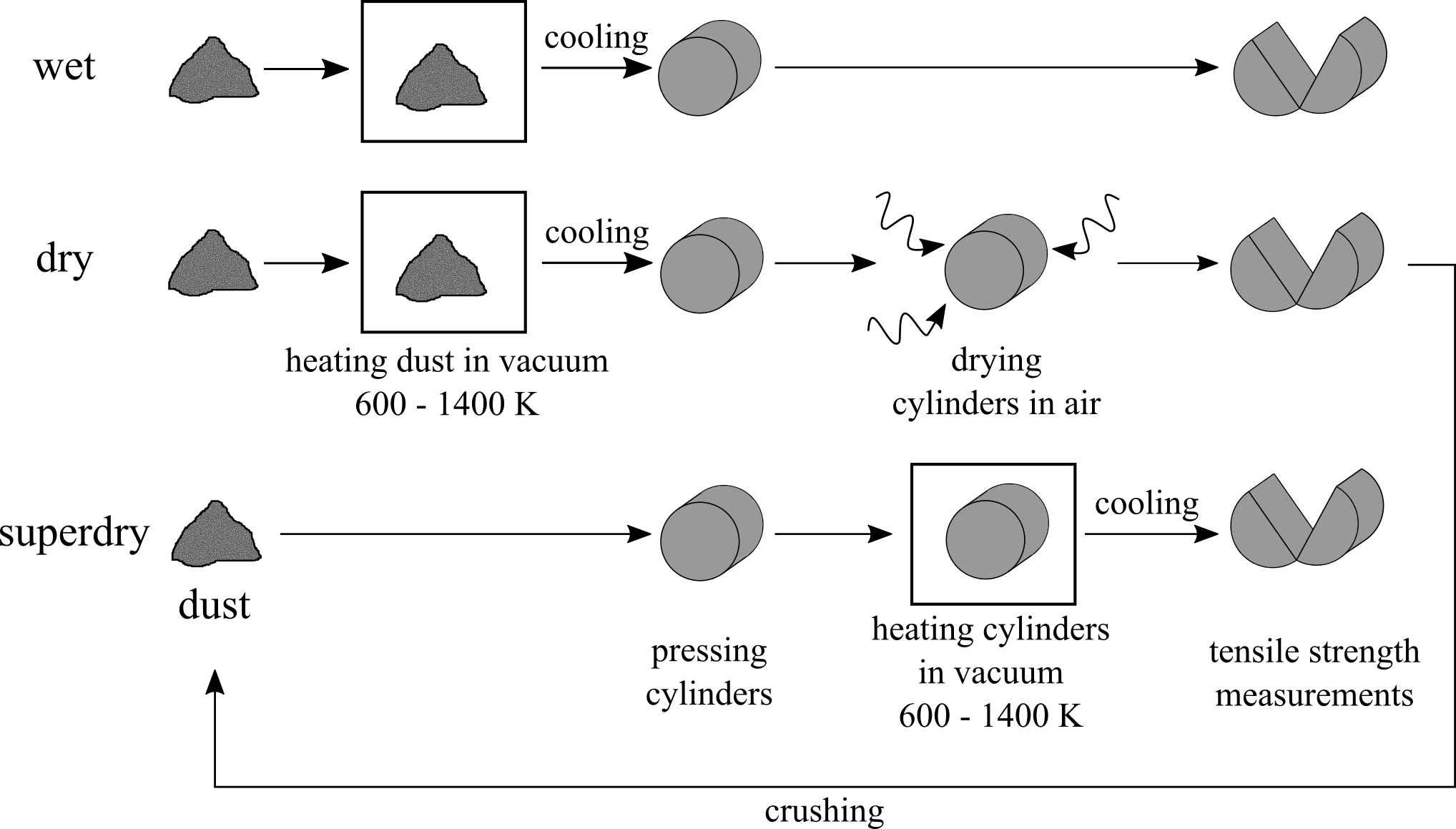}
        \caption{\label{fig.schema_prep} Illustration of the sample preparations and measurement sequences. After tensile strength measurements, the cylinders were crushed, and the sequence of each sample was repeated with a higher heating temperature.}
\end{figure}

We used the same chondritic material in this study as in \citet{Bogdan2020b}, that is, the remaining mass of an L4/5 chondrite. To study the effect of surface water on the sticking properties, the dust was pressed into cylinders and dried for 24 h in normal atmosphere at 523 K. The data set of these samples is called "dry" in this work. As this process could lead to oxidation of the iron contents, we also used a second sample set, designated "super dry": We followed the same procedure as in our previous work, but the material was pressed into cylinders before heating in an evacuated oven ($\mathrm{10^{-3} - 10^{-2} \, mbar}$) at temperatures between 600 and 1400\,K.
All different processing steps of the chondritic dust from preparations over heating and drying to measurements of the tensile strength are shown in fig. \ref{fig.schema_prep}.

To test the idea of the water effect on minerals other than amorphous silica and to develop our procedure of removing water, we also used commercial basalt samples. This is more readily available in large amounts. In both cases, the material was milled to micrometre size in a first step. The wet and dry samples were processed as shown in fig. \ref{fig.schema_prep}, but without the first heating step.

\subsection{Basalt sample, wet and dry}

\begin{figure}
        \includegraphics[width=\columnwidth]{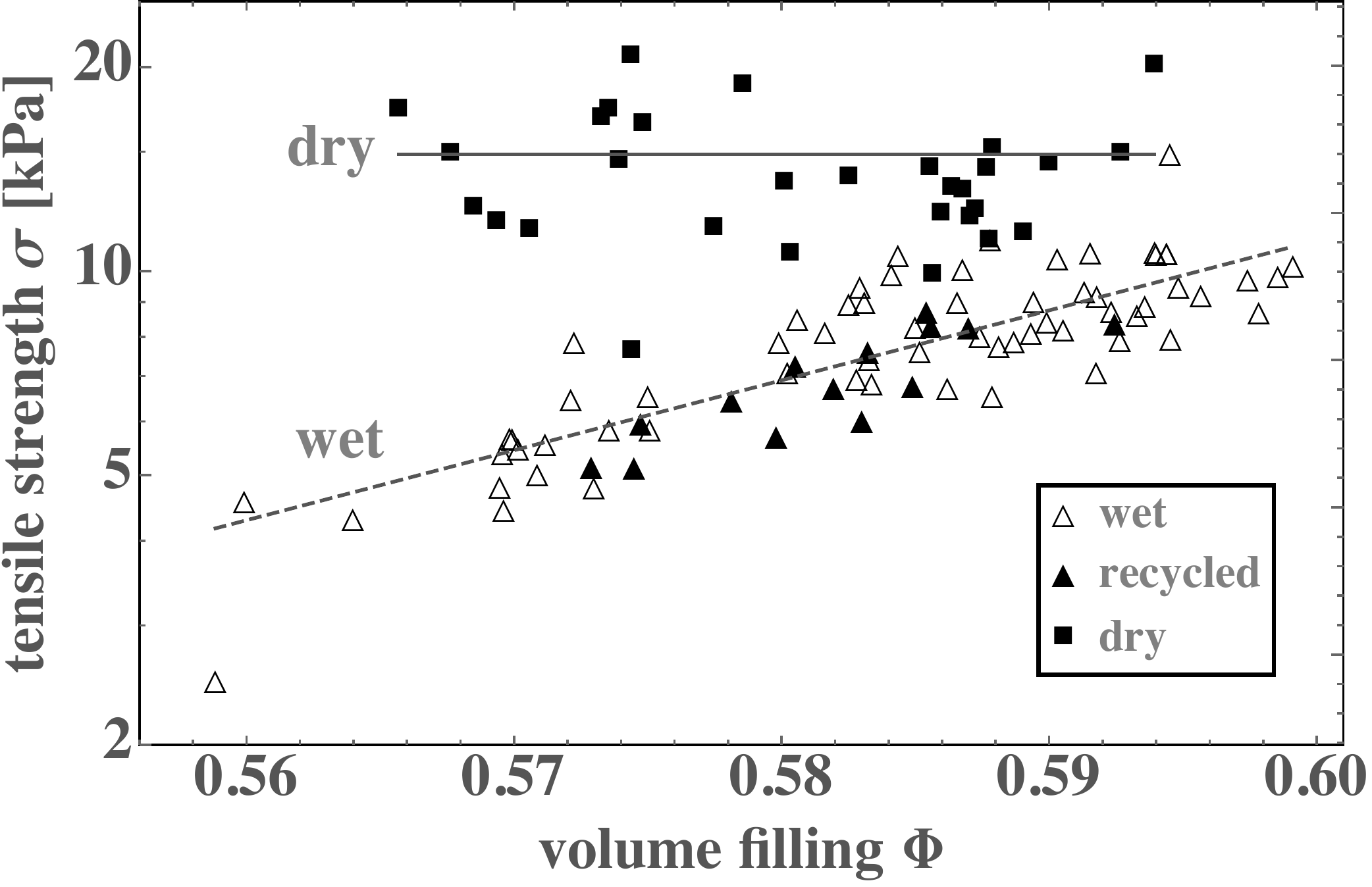}
        \caption{\label{fig.moving} Tensile strength over volume filling factor of the basaltic samples. 
        The recycled sample was dried, crushed, and reprepared before measuring. The dashed line is a power law for the wet samples, and the continuous line is an average constant for the dry samples.}
\end{figure}

We measured the splitting tensile strength on cylinders of compressed dust, which break into halves if subjected to a load on the mantle (Brazilian test).
Previous works on tensile strength regularly find a power-law dependence on the volume-filling factor \citep{Meisner2012,  Steinpilz2019, Bischoff2020}, and so did we in paper\,I \citep{Bogdan2020b}. These power laws agree with models by \citet{Rumpf1970} and can be traced back to variations in grain number density associated with the filling factor $\Phi$, changes in contact forces $F,$ and variations in the number of contacts per grain in a dust aggregate (cylinder) $N$ as outlined in paper\,I, or  
\begin{equation}
\sigma = \left(\frac{1}{\pi} \right) \left( \frac{1}{d^2} \right) \left(\frac{\Phi}{\Phi_0}\right)^{a_1+a_2+1} N_0 F_0 \Phi_0.
\label{rumpf4}
\end{equation}
We note that we had an additional factor of 9/8 from previous work in this equation in paper\,I, which we omit now. However, because this is close to 1, it does not imply significant changes and does not affect the comparisons later.
Eq. \ref{rumpf4} assumes power laws for all three quantities $N, F,\text{and } \Phi,$ with the index 0 applying to a given volume-filling factor $\Phi_0$ , and powers $a_1$ and $a_2$ applying to the power laws of the force and number of contacts, respectively.

We would like to point out that contact forces refer to the average force between constituent grains. This includes pull-off forces, but also sliding and rolling forces \citep{Omura2017}. Which one dominates depends on the sample. As grains shift with different filling factors, this average force can also change with filling factor.

With this in mind, we measured the tensile strength of basalt samples, as shown in fig.\,\ref{fig.moving}. The figure displays wet and dry samples. Wet samples were prepared by pressing a cylinder at ambient conditions prior to measurement. For the dry samples, the cylinders were heated 24 hours at 520\,K before measuring the tensile strength. This removes water from the contacts presumably down to a monolayer.
The wet samples show the typical power law on filling factor that we expected. However, the dried samples behave differently. 

We first note that the tensile strength is significantly higher for the dry samples. Dry samples measured immediately after drying or measured after spending one week under normal laboratory conditions did not show a significant difference in this behaviour. 
With respect to our first paper \citep{Bogdan2020b}, it might be tempting to attribute this increase in tensile strength to changes in composition upon drying (moderate heating) or changes in grain size. However, on the other hand, the strong increase corresponds to the findings by \citet{Steinpilz2019} for pure $\rm SiO_2$ grains, which neither changed size nor composition upon drying. \citet{Steinpilz2019} explained the increase in tensile strength by the loss of surface water. 
A strong indication that the loss of water also drives the increase here comes from recycled dried cylinders. For a number of cases, we crushed the dried samples after measurement of the tensile strength and reassembled the dust into new cylinders. Tensile strength measurements of these recycled samples are shown as filled triangles in fig. \ref{fig.moving}. They behaved exactly like wet cylinders, again with the same low tensile strength. Water loss therefore generally explains an increase in tensile strength.

Most surprisingly, however, the dependence on the filling factor changes strongly between wet and dry samples. Essentially, dried samples no longer show a dependence on filling factor.   
The total power index in eq. \ref{rumpf4} clearly decreases strongly when water is removed. 
Our interpretation of this is as follows.

In the limited range of filling factors, the direct $\Phi$-dependence (power 1) is of minor importance. Although grains might shift, the general contact forces $F$ should also vary little in the limited range because this contact force is an average of the forces acting between constituent grains. These include pull-off forces of different curvature contacts for irregular grains, sliding, rolling, and twisting forces. A systematic change in contact forces might be relevant when grains change size and shape, such as the chondritic dust does close to 1400\,K \citep{Bogdan2020b}, but this is of no relevance for the basalt samples. This leaves the number of contacts per grain, which no longer depends on the filling factor for the dried samples. 

\begin{figure}
\includegraphics[width=\columnwidth]{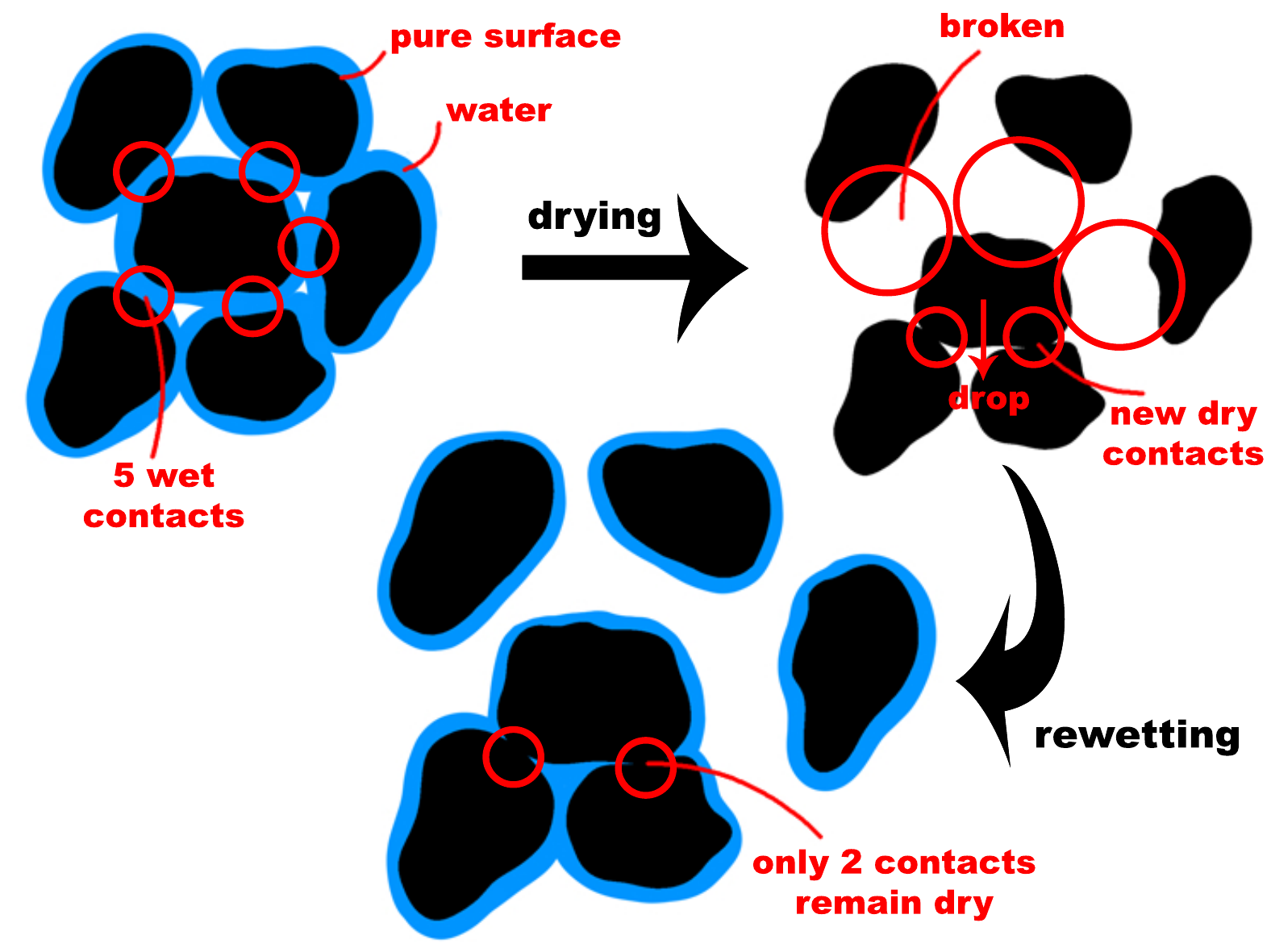}
        \caption{\label{fig:dryingcontacts} As a sample contracts upon drying, a smaller number of new and dry contacts is established between the grains. Even rewetting under laboratory conditions will not increase the number of contacts again, and water will not diffuse back into the contact area. The most stable configurations in which no more rotations are possible have N=3 contacts (N=2 in two dimensions).}
\end{figure}

This is plausible, however, as visualized in fig. \ref{fig:dryingcontacts}.
Dry samples are not prepared with dry dust, that is, they are not pressed into cylinders from dried dust. We currently cannot do this with confidence, but we would expect the typical increase of tensile strength with filling factor then. In our case, samples are only dried after they are already cylinders (also see sequence sketch of preparations in fig. \ref{fig.schema_prep}). If a cylinder is dried, several layers of water are removed from a contact. If grains are not free to move, contacts are eventually broken. A too simple but probably illustrative comparison might be a mud surface that cracks upon drying. In other words, if we were to fix all grain positions and removed all water from the surfaces, the remaining solid grains would no longer be in contact. Particles then rearrange slightly to make contact again. With enough cohesion and without external pressure, one contact per grain could already suffice. With another grain on top, each grain would have two contacts. Without cohesion, a stable configuration of a grain on a surface would have three contacts, and likely one on top or a total of four contacts. Because our grains are highly cohesive but because two is only the minimum, we assumed $N=3$ as number of contacts per grain. We note that this is only a model. It does not have to be $N=3,$ but a constant contact number would agree with the data as the sample shrinks upon loss of water. It should be kept in mind that this induces a systematic uncertainty.

As a reminder to paper\,I, we calculated the effective surface energies according to
\begin{equation}
    \gamma_e = 1.3 \cdot d \cdot \frac{\sigma_0}{N_0 \Phi_0},
    \label{gammedef}
    \end{equation}
with grain size $d$.  We note that the prefactor changed from 1.2 to 1.3 as we removed the factor 9/8 in eq. \ref{rumpf4}.

For the wet basalt sample, this yields $\gamma_e = 0.02 \, \rm{J m^{-2}}$ with $d = 7.8 \, \rm{\mu m}$, $\sigma_0 = 6.90 \, \rm{kPa}$, $N_0 = 6.5$ and $\Phi_0 = 0.58$. For the dry samples, it is $\gamma_e = 0.08 \, \rm{J m^{-2}}$, roughly a factor 4 higher using $N_0 = 3$ and $\sigma_0 = 13.9 \, \rm{kPa}$.

\subsection{Wet chondritic sample}

To deduce the effect of the water content on the surface of chondritic grains, we used the measurements of paper I \citep{Bogdan2020b} as reference for the wet chondritic sample. To maintain consistency between dry and wet samples, we merely adapted the volume filling to account for the water content within the cylinders. The water content is derived from the difference in mass of the cylinders before and after drying. 

\begin{figure}
        \includegraphics[width=\columnwidth]{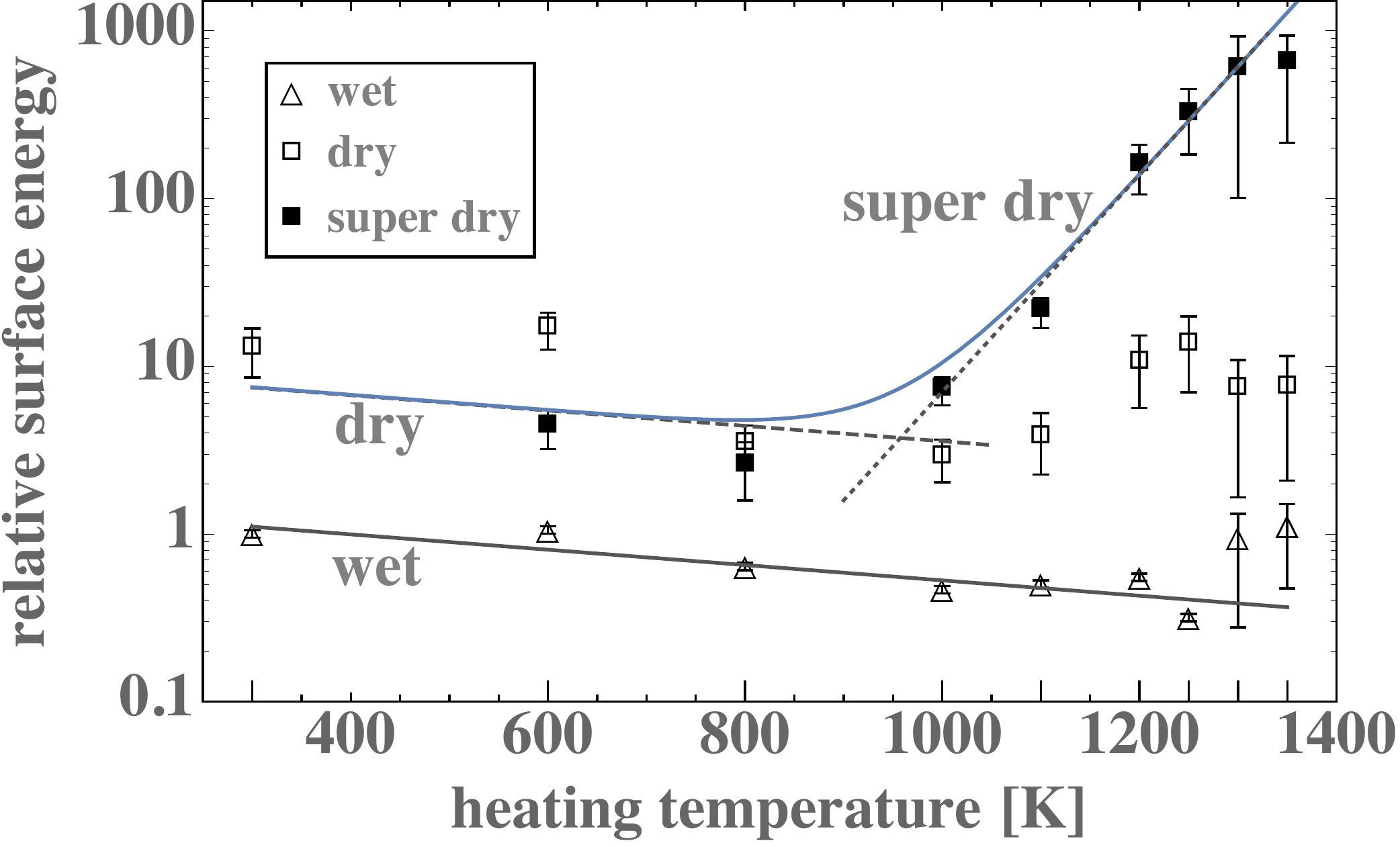}
        \caption{\label{fig.gammafactors}Relative surface energy ($\gamma_e/\gamma_{e_{\rm{w}}}(300\,\rm K)$) for the wet (triangles), dry (open squares) and super dry (filled squares) chondritic sample; Continuous line: exponential trend using data of wet sample up to 1250\,K; Dashed line to dry data points: exponential trend to the data up to 1000\,K including super dry data up to 800\,K with same inclination as for wet samples; Dotted line to super dry data: exponential trend based on data up to 1300\,K; Blue line: our model for the overall effective surface energy dependence on temperature.}
\end{figure}

{As in paper\,I, we used 
 $\gamma_e = 0.07 $ $ \rm   J m^{-2}$ at room temperature (300 K).} To highlight the dependences, we again give a relative surface energy by relating all values to the value of wet dust at room temperature or $\gamma_e/\gamma_{e_{\rm{w}}}(300\,\rm K)$. For the wet sample, this gives the same data set as in paper\,I, shown here in fig.\,\ref{fig.gammafactors} (lowest data set).

\subsection{Dry chondritic sample}

\begin{figure}
        \includegraphics[width=\columnwidth]{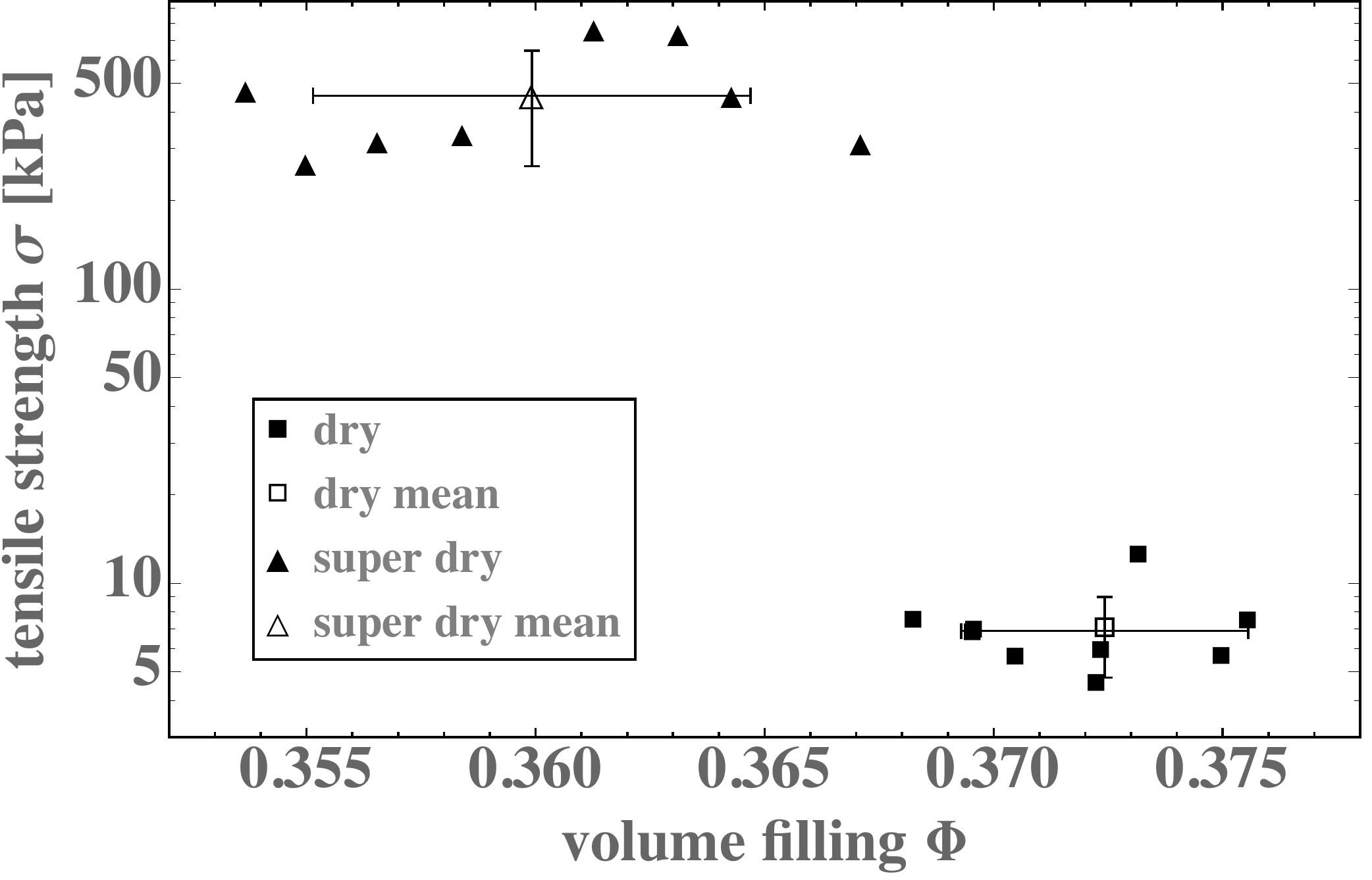}
        \caption{\label{fig.gammafactorsexample}Example of individual tensile strengths over volume-filling factors of the data of dry and super-dry chondritic samples heated at 1300\,K. Also shown are the average values and their standard deviations. The average values are used to calculate the surface energy in eq. \ref{gammedef}.}
\end{figure}

Based on the results from the basalt sample, drying for 
24\,h at 523\,K seemed reasonable to remove but a monolayer of water. We therefore applied the same method to the tempered chondritic sample.
We milled the remaining mass of the same chondrite used in paper\,I to micrometre dust of the same size as in \citet{Bogdan2020b}, and again tempered it for 1\,h before each measurement in a temperature range from 600\,K to 1400\,K in vacuum, just like in paper\,I.
In addition, for each heating temperature, the cylinders were dried before measuring. 
The relative surface energy data of the dried samples is also shown in 
fig.\,\ref{fig.gammafactors}. As outlined above, we used $N_0 = 3$ for the calculations of the surface energy according to eq. \ref{gammedef}. As seen in fig. \ref{fig.gammafactorsexample} for measurements of dust treated at 1300\,K as a typical example, the volume-filling factors only have a small spread for a given data set. Therefore we used the average values of $\Phi_0$ for the calculations of the surface energy. Up to about 1000\,K, the data trace the wet sample, but shifted by a factor 10. The value at 600\,K is somewhat high, but we only use simple linear trends in the figure or exponential dependences of the data.   That is to say, the compositional changes underlying the water layers are still visible in the same way, being about a factor 5 in reduction from 300\,K to 1250\,K if we ignore the measurements beyond 1000\,K.

The data at 1200\,K and higher again reach the value of dry chondritic material at room temperature. 
\begin{figure}
        \includegraphics[width=\columnwidth]{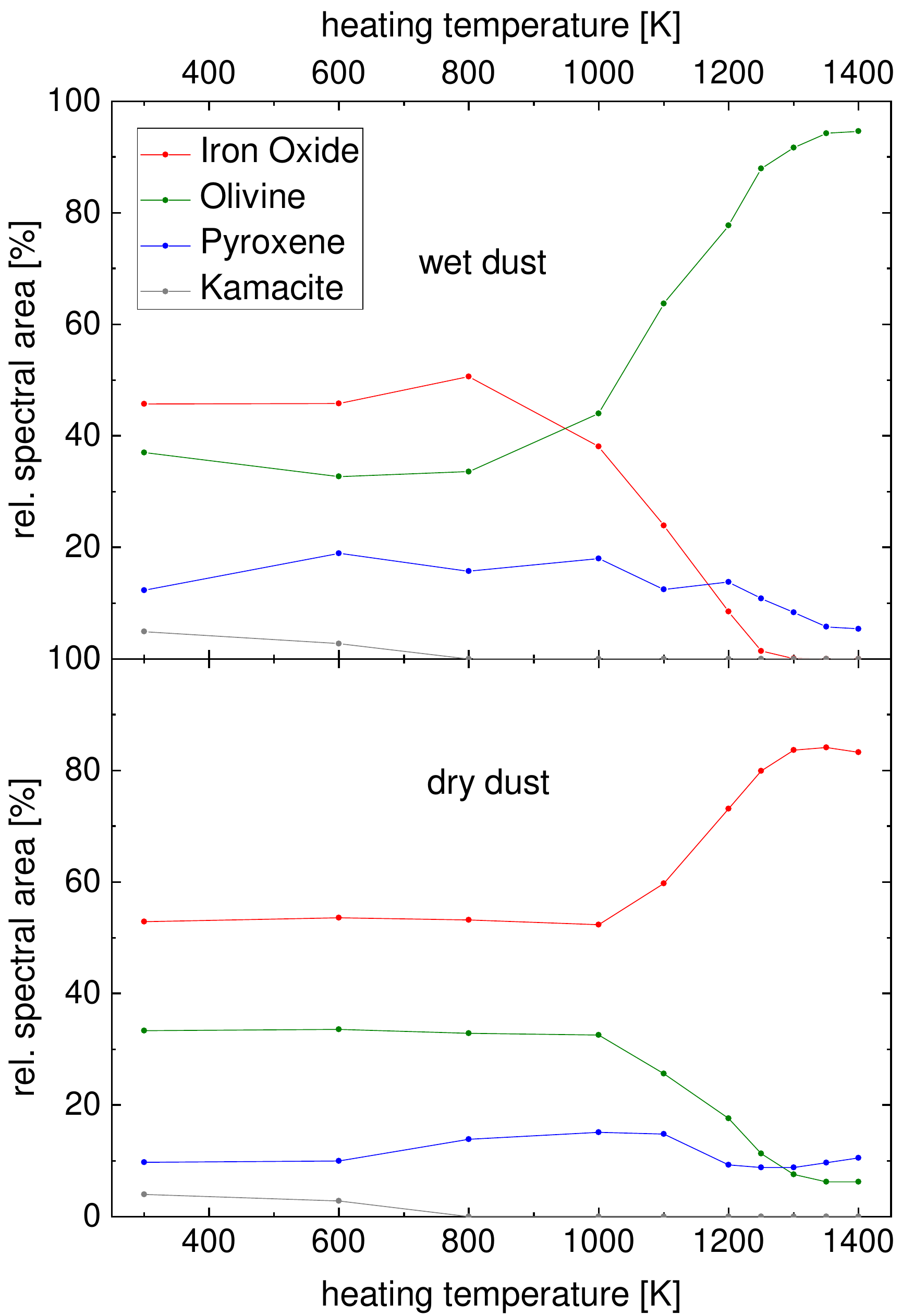}
        \caption{\label{fig.relative_spectral_areas} Relative spectral areas of wet (top image) and dried powder (bottom image) used to calculate the density of solids. We note the drastic changes in composition for heating temperatures $\mathrm{T_H}$\,>\,1000\,K.}
\end{figure}
This can be explained by a change in composition above 1000\,K as traced by Mössbauer spectroscopy and shown in fig.\,\ref{fig.relative_spectral_areas}. The relative spectral areas of the minerals found in the chondrite were deduced by Mössbauer spectra measured at room temperature. To ensure a detailed identification of each component, they were compared to low-temperature spectra. For a more elaborate explanation of this procedure, we refer to \citet{Bogdan2020b}. 
There we found a decline in iron oxides at higher temperatures and an increase in the olivine fraction (top in fig.\,\ref{fig.relative_spectral_areas}).
Here, we have to note that we have an additional heating process between tempering in vacuum and tensile strength measurement, namely the drying at 523\,K for 24\,h. This obviously has no effect on the composition for tempering up to 1000\,K as seen in the bottom image of fig.\,\ref{fig.relative_spectral_areas}, but at higher temperatures, it removes iron from the olivines and again generates iron oxides. This agrees with the measured surface energies as iron oxide now again dominates the sticking.
However, for comparison, we therefore did not consider these high-temperature data for dried dust here.

\subsection{Super-dry chondritic sample}

To avoid the formation of iron oxides, drying would have required heating in an inert atmosphere or heating under vacuum. We considered the above series of dry measurements as consistent enough, however, to not proceed the measurements of dry samples. Following this idea, and as we already temper in vacuum, the same setup might be used for tempering and drying at the same time if we already temper cylinders. This is of special interest because \citet{Kimura2015}, for instance, not only specified that the surface energy should increase by a factor 10 with only a monolayer or a silanol surface, but should increase by another factor if the remaining water is removed, in their case, leaving a siloxane surface. We would therefore expect an increase in surface energy for these super-dry surfaces.

Our super-dry data are also shown in fig.\,\ref{fig.gammafactors} to agree with the dry sample up to 800\,K. After an exposure to 900\,K, the surface energy starts to deviate from the dry sample, that is, it increases exponentially with temperature until it reaches another factor 100 at about 1300\,K.
Chondritic dust at high temperatures is clearly very sticky. This behaviour is not explained by temperature-induced phase transitions as the composition of the wet and super-dry sample are very similar. The top panel of fig.\,\ref{fig.relative_spectral_areas} exemplarily shows the temperature-dependent composition of wet samples.

\section{Conclusion}

The growth of dust aggregates in protoplanetary discs strongly and sensitively depends on the sticking forces between grains \citep{Pinilla2020}. If grains can grow large enough, the probability that they might evolve into planetesimals might increase.

In paper\,I we found that wet chondritic matter with some layers of water on the surface reduces its effective surface energy by a factor of 5 from room temperature to about 1300\,K due to compositional changes. Here we find that dry samples show the same trend of reduction, and a factor of 5 still holds. 

In addition, water reduces the sticking forces by a factor of 10 compared to dry samples or only a monolayer of water.  This confirms and extends the findings by \citet{Kimura2015} and \citet{Steinpilz2019} in the sense that the factor 10 they reported for wet and dry amorphous silica is also valid for a more complex and realistic mineral mix of protoplanetary discs. 

 The boost in surface energy for super-dry or high-temperature dust has not been considered before for dust in protoplanetary discs. The compilation of surface energies by \citet{Kimura2015} indicates an increase as all water is gone on amorphous silica. Our data show, however, that the increase for the chondrite is even stronger than might have been expected. 

The overall effective surface energy dependence on temperature might well be modelled by a combination of compositional changes and water effects or
\begin{equation}
    \gamma_e = \gamma_{e_{\rm{w}}} \left( f_{\rm{w}} \, e^{ - \frac{T-300 \, \rm{K}}{T_{\rm{w}}}} + f_{\rm{sd}} \, e^{\frac{T-300 \, \rm{K}}{T_{\rm{sd}}}} \right) .
\end{equation}
The first part has a reference value for the wet dust at 300\,K $\gamma_{e_{\rm{w}}}$, a factor relating dry samples in discs to wet samples in the laboratory $f_{\rm{w}}$, and a characteristic temperature $T_{\rm{w}}$. The second term accounts for the super-dry incline. It relates to the wet samples at 300\,K by a factor $f_{\rm{sd}}$ and has a characteristic temperature $T_{\rm{sd}}$.
For $f_{\rm{w}}$ , the best fit to the data is $f_{\rm{w}} = 7.5$ or roughly $f_{\rm{w}} \sim 10$, keeping in mind that the few data points and observed variations only restrict this ratio to a factor 2.
For $f_{\rm{sd}}$ , we obtain $f_{\rm sd} = 0.0002$, which is orders of magnitudes lower. This shows that it is not important for room temperature.
The characteristic temperatures are $T_{\rm{w}}=950.06 \,\rm{K}$ and $T_{\rm{sd}}=67.16 \,\rm{K}$.
The reference value of $\gamma_{\rm{e}_{\rm{w}}} = \rm 0.07 \, J/m^2$ has to be used with care if absolute values are required because this is an effective value based on some assumptions (see paper\,I). If better values are determined or other base values, for example, for amorphous silicates are known, this might just be substituted while keeping the water and temperature dependence. 

\begin{figure}
        \includegraphics[width=\columnwidth]{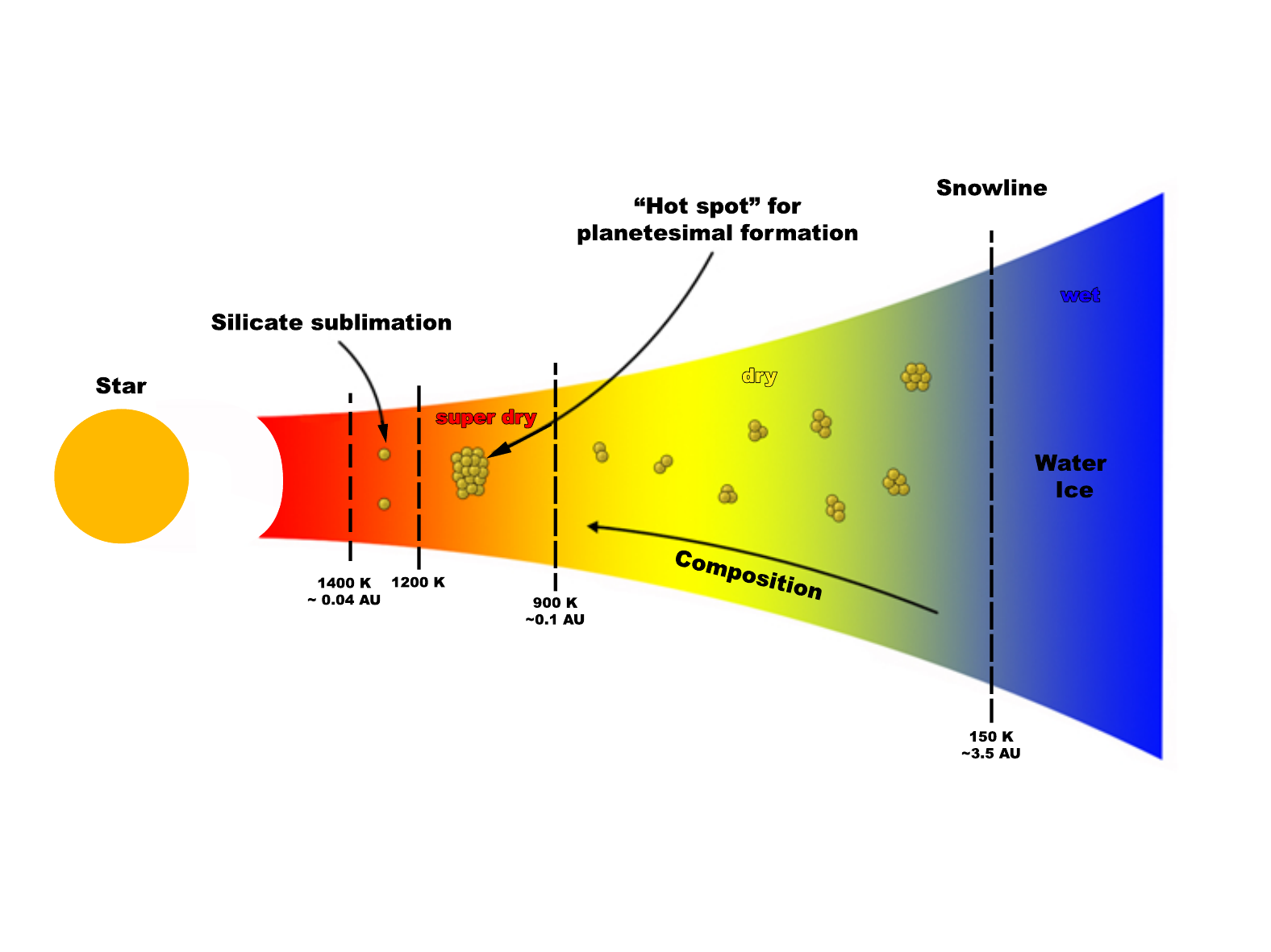}
        \caption{\label{fig:diskmodel} Proposed scenario for protoplanetary discs: Removal of water in the warm inner part of discs leads to stronger contacts, which are possible hot spots for planetesimal formation. Temperatures and distances refer to minimum mass nebula.
        }
\end{figure}

This relation is valid in a temperature range between 300\,K and 1300\,K. Between 1300\,K and 1400\,K, grain sizes increase, leading to an instability of aggregates and therefore strongly making growth challenging. Beyond 1400\,K no classical collisional growth is possible (see paper\,I). 
These concepts are visualized in fig. \ref{fig:diskmodel}.

In any case, an exposure to a narrow range of temperatures from 1000\,K to 1300\,K leads to an increase in sticking of up to a factor 100, rendering this region interesting for planetesimal formation that can likely be considered a sweet spot for aggregate growth.

\section*{Acknowledgements}

This work is supported by DFG grants WU 321/18-1 and WE 2623/19-1.

\bibliographystyle{aa}
\bibliography{bibbi} 

\label{lastpage}
\end{document}